\documentstyle[psfig]{mn}

\def\gs{\mathrel{\raise0.35ex\hbox{$\scriptstyle >$}\kern-0.6em
\lower0.40ex\hbox{{$\scriptstyle \sim$}}}}
\def\ls{\mathrel{\raise0.35ex\hbox{$\scriptstyle <$}\kern-0.6em
\lower0.40ex\hbox{{$\scriptstyle \sim$}}}}
\def\ls{\mathrel{\hbox{\rlap{\hbox{\lower4pt\hbox{$\sim$}}}\hbox{$<$}}}}
\def\gs{\mathrel{\hbox{\rlap{\hbox{\lower4pt\hbox{$\sim$}}}\hbox{$>$}}}}

\def\mnras {{\sc MNRAS}}

\def\aj {AJ}

\title[The Large Peculiar Velocity of the cD Galaxy in Abell 3653]
      {The Large Peculiar Velocity of the cD Galaxy in Abell 3653}
\author[Pimbblet, Roseboom \& Doyle]
       {Kevin A.\ Pimbblet$^{1,2}$, Isaac G.\ Roseboom$^{1}$, 
        Marianne T.\ Doyle$^{1}$ 
        \vspace*{1mm}\\
        $^{1}$Department of Physics, University of Queensland, Brisbane,
        4072 Queensland, Australia\\
        $^{2}$email: pimbblet@physics.uq.edu.au}

\date{\fbox{\sc Draft: \today\ --- Do Not Distribute}}

\pagerange{000--000}

\begin{document}

\maketitle

\begin{abstract}
We present a catalogue of galaxies in Abell~3653 from observations
made with the 2dF spectrograph at the Anglo-Australian Telescope.
Of the 391 objects observed, we find 111 are bone-fide members
of Abell~3653.  We show that the cluster has a velocity of
$cz = 32214 \pm 83$ kms$^{-1}$
($z=0.10738 \pm 0.00027$), with a velocity dispersion 
typical of rich, massive clusters 
of $\sigma_{cz} = 880^{+66}_{-54}$.
We find that the cD galaxy has a peculiar velocity of
$683 \pm 96$ kms$^{-1}$ in the cluster restframe -- some
$7\sigma$ away from the mean cluster velocity, making it 
one of the largest and most significant 
peculiar velocities found for a cD galaxy to date.  
We investigate the cluster for signs of substructure,
but do not find any significant groupings on any length
scale.  We consider the implications of our findings on 
cD formation theories. 

\end{abstract}

\begin{keywords}
surveys -- catalogues -- galaxies: clusters: individual: Abell 3653 -- 
galaxies: elliptical and lenticular, cD --
galaxies: kinematics and dynamics
\end{keywords}

\section{Introduction}

Clusters of galaxies permit the study of a large number of
galaxies ($\sim 10^3$) at a common distance (likely co-eval) -- 
this makes them ideal laboratories for studying galaxy evolution.
They generally feature predominantly early-type 
(elliptical and lenticular) galaxies in their core regions,
often with a central cD galaxy in residence.
cD galaxies are an extreme example of giant ellipticals:
they are surrounded by a low surface brightness envelope and
are only ever found in groups or clusters: never in isolation
in field (i.e.\ low density) environs. 
The formation mechanism(s) for cD galaxies has been hotly
debated in the literature for some time.

The formation of a central cluster galaxy may be explicitly
associated with the 
post-virialization accretion of smaller mass galaxies by the 
central galaxy.  This process, devised by Ostriker \& Tremaine
(1975), is aptly known as galactic cannibalism (see also
Nipoti et al.\ 2004; 2003;
Garijo, Athanassoula, \& Garcia-Gomez 1997; 
Blakeslee \& Tonry 1992; 
Capelato et al.\ 1985;
Duncan, Farouki, \& Shapiro 1983;
Hausman \& Ostriker 1978).
The theory predicts that we must be able to
observe the process happening: multiple cores of
cD and D galaxies in clusters should be visible.
Indeed, it is the case that many cD galaxies are found to have 
multiple nuclei (e.g.\ Yamada et al.\ 2002;
Gregorini et al.\ 1994;
Blakeslee \& Tonry 1992;
Merrifield \& Kent 1991; 
Lauer 1988).

It is possible at higher redshifts to see the progenitors 
of cD galaxies as multiple cluster galaxies coming together.
Yamada et al.\ (2002) shows that the brightest cluster galaxy
in a cluster at $z=1.26$ is composed of two distinct sub-units
that are likely to fully merge on a timescale of $10^{8}$ years.
Their result is broadly in agreement with numerical predictions
that show central galaxies grow through repeated mergers of smaller
mass galaxies (e.g.\ Dubinski 1998; Garijo et al.\ 1997).

However, for the cD galaxy to attain enough mass and luminosity
to account for these observations, 
it must be preferentially situated at the cluster centre where
the cannibalistic accretion process is at peak efficiency 
(Barnes \& Hernquist 1992; Tremaine 1990).
Indeed, in such a post-virialization formation model of cD galaxies,
dynamical friction between the inter-galactic medium and the 
cluster galaxies
will cause the larger cluster 
galaxies to fall to the centre of a cluster
and eventually merge with the cD galaxy that is already there.

Galactic cannibalism cannot be such a strong driving mechanism, though.
Lauer (1988) shows that cannibalism cannot account for the large
observed luminosities of cD galaxies.
Another problem that Merritt (1985) points out with this scenario
is that galaxy halos will be disrupted by the tidal field of
a cluster.  This leads directly to the dynamical friction timescale 
being increased and hence the effect of galactic cannibalism
will be decreased significantly.
Merritt (1984) and Smith et al.\ (1985) 
also suggest that the majority of 
multiple-nucleus systems are transient phenomena, and are 
not galaxies in the process of merging.  

As an alternative to the post-virialization formation of cD 
galaxies, Merritt (1985; 1984) suggests that they could form before
or during the virialization of rich clusters.  Moreover, cD galaxies 
must also form at roughly the dynamical centre of the cluster
to avoid having its outer envelope truncated by tidal forces.
Many observations confirm that cD galaxies are located at the
dynamic centres of galaxy clusters 
(e.g.\ Oegerle \& Hill 2001; 
Quintana \& Lawrie 1982).
However, if clusters grow hierarchically through repeated mergers
of sub-clusters and galaxy groups, then presumably
the cD galaxy would have already been formed in one of these
sub-clusters before drifting to the centre of the cluster.

This would yield two observable 
`smoking-guns'.  The first would be the
presence of substructure in a cluster meaning that the cluster
is a dynamically young entity.  The second would be a high
peculiar velocity of the cD galaxy with respect to the parent
cluster.  We note that these two quantities are related.
Oegerle \& Hill (2001) show that substructure can 
account for, and perhaps even drive, cD peculiar velocities.
Indeed, a cD galaxy may even be considered as a road-sign
to substructure (Zabludoff et al.\ 1993; 
Beers \& Geller 1983).

Malumuth (1992) explores if it is possible to form cD galaxies with 
large residual peculiar velocities in N-body simulations of cluster
formation and growth.  These simulations not only include the effect
of dynamical friction and two body relaxation, but also comprise 
the role of galaxy-galaxy stripping collisions and mergers.
Within $\sim 10^{10}$ years, Malumuth (1992) shows that 
the cD galaxies in the simulations have had any significant
peculiar velocities removed and they have arrived 
at the centre of cluster through dynamical friction.

However, there is a growing body of evidence that a non-negligible
fraction of clusters possess cD galaxies with significant
peculiar velocities (Oegerle \& Hill 2001 and references therein). 
Hill et al.\ (1988) report a cD peculiar velocity in
Abell~1795 of 365 kms$^{-1}$. 
In their observations of Abell~2670,
Sharples, Ellis \& Gray (1988) find the cD galaxy has
a peculiar velocity that is 439 kms$^{-1}$ -- 
$3.7\sigma$ -- away from the cluster mean.  
Later, Bird (1994; see also Bird 1993; Oegerle \& Hill 2001) 
showed that the
cD peculiar velocity may be intimately tied to the
presence of substructure in Abell~2670.  Such an effect
lends support to the picture of clusters growing
hierarchically through repeated mergers. 
For the simulations of Malumuth (1992), it suggests that 
the formation of cD galaxies after the virialization of the
cluster cannot account for such cD peculiar velocities.
Therefore some or all of the following could be true: 
(1) clusters that form cD galaxies are necessarily young; 
(2) cD galaxies are a relatively recent occurrence;
(3) clusters of galaxies are not virialized\footnote{i.e.\ the 
mean velocity of the galaxies does not represent the 
cluster velocity.};
(4) cD galaxies do not form in their present environs, rather
they have been accreted from elsewhere;
(5) real Universe dynamical friction is less efficient than
simulated.
In this work, we present evidence for one of the largest 
and most significant cD peculiar velocities found to date
in Abell~3653.

The format of this work is as follows.  In Section~2, we describe
the selection, observations and reduction of our 2dF dataset. 
In Section~3, we explore the redshift structure of Abell~3653
and derive its velocity and velocity dispersion.
In Section~4, we closely examine the cD galaxy of Abell~3653
and its highly peculiar velocity.
In Section~5, we probe the cluster's morphology for any sign of
substructure.  We discuss our findings in Section~6 before
summarizing the results in Section~7.
The full spectroscopic catalogue of our observation is presented
in Appendix~A.
Throughout this work we use a cosmological concordance model with
values of $H_0 = 70$ kms$^{-1}$Mpc$^{-1}$, $\Omega_M = 0.3$ and
$\Omega_\Lambda = 0.7$.

\section{Data and Data Reduction}

It is essential to note here that 
the original aim of our observations was to confirm or
deny the presence of a dark galaxy candidate in the region
(Doyle et al.\ 2005).  Such an object is one that is visible 
at radio wavelengths (i.e.\ HI), but has no obvious optical
counterpart.  The dark galaxy candidate was later eliminated
by other follow-up observations (see Doyle et al.\ 2005 for
more detail).  However, the 2dF data gathered for that aim were
approximately centered upon Abell~3653 and, here, we describe 
this unique dataset in full.

\subsection{Observations}

Our observations for this work come from 2dF spectroscopy and
are summarized in Table~\ref{tab:obs} 
whilst we present our final catalogue in Appendix~A.
These observations are entirely performed in service mode on the AAT,
meaning that we have a mixture of gratings (300B and 270R),
and physical conditions.  Regardless, all of our observations are
approximately centred on [OIII]$\lambda5007$\AA \ which gives us 
coverage of major spectral lines such as H$\alpha$, H$\beta$, H$\gamma$,
H$\delta$, and (for many) 
Ca K \& H at the median redshift of Abell 3653 ($z\sim0.1$)
-- i.e.\ more than adequate to provide plentiful features 
to measure redshifts with.

%
%
\begin{table*}
\begin{center}
\caption{Summary of 2dF observations.  Note that the majority 
of observations made on
28 July 2005 are re-observed in subsequent configurations.
Each configuration, N(config), refers to one 2dF CCD worth of
observations, equivalent to a total 200 fibres.  N(obs) is the
number of fibres out of the 200 possible fibres that are allocated
to our spectroscopic targets; the remainder are divided between
sky fibres and unused (or) dead fibres.
\hfil}
\begin{tabular}{lcccccc}
\noalign{\medskip}
\hline
Observation & N(config) & Integration & Grating & Seeing   & N(obs) \\
Date        &           & Time (s)    &         & (arcsec) &         \\
\hline
28 July 2005    & 1 & $1\times 1800$ & 300B & 1.6--2.2 & 183      \\
 9 August 2005  & 2 & $3\times 1800$ & 270R & 1.2--1.8 & 134, 130 \\
10 August 2005  & 1 & $3\times 1800$ & 270R & 1.7--1.9 & 160      \\
\hline
\noalign{\smallskip}
\end{tabular}
  \label{tab:obs}
\end{center}
\end{table*}

We note that 2dF has fibres that range in size from
2.16'' diameter on the optical axis to 1.99'' at the edge of
a two degree field.  At the adopted cluster distance,
1'' corresponds to 1.941 kpc, giving an aperture size of 4.19--3.86 kpc 
depending upon where a fibre is placed.  We note that the cD galaxy
is near the centre of the 2dF field, but nevertheless, the
measured velocity for the cD may be biassed if there are 
multiple nuclei present.  

The galaxies chosen for spectroscopic observation are derived from
the APM catalogue (e.g.\ Maddox et al.\ 1990; see also
www.ast.cam.ac.uk/$\sim$mike/apmcat/).  Our broad 
aim is to sample this
region of the sky without bias to colour or galaxy type,
down to a limiting magnitude of $R\approx18$ (roughly corresponding
to $M^{\star}+1.5$ at the redshift of Abell~3653) 
in order to probe sufficiently deep
in the luminosity function (i.e.\ to eliminate all obvious
bright galaxies that might have been optical counterparts to
the dark galaxy candidate in the region; Doyle et al.\ 2005). 
Therefore we 
select all objects within the 60 arcmin of the nominal cluster
centre ($\alpha =$ 19 52 37, $\delta =$ -52 01 14; NASA Extra-galactic 
Database\footnote{The NASA/IPAC Extragalactic Database (NED) is operated
by the Jet Propulsion Laboratory, California Institute of Technology, 
under contract with the National Aeronautics and Space Administration.}).  
Further, we made no attempt to exclude stars from our
selection so that we would not bias ourselves to excluding more compact
galaxies; akin to the approach adopted by Drinkwater et al.\ (2003) in
the nearby Fornax and Virgo clusters.  As noted by Drinkwater, this
is an observationally expensive technique that in our case 
results in almost as many stars being observed as galaxies.
The APM positional accuracy (Maddox et al.\ 1990) is 
better than 0.3'' for all galaxies -- sufficient
for 2dF observations and is also the same approach used by Colless et
al.\ (2001) for the Galaxy Redshift Survey.

\begin{figure}
\centerline{\psfig{file=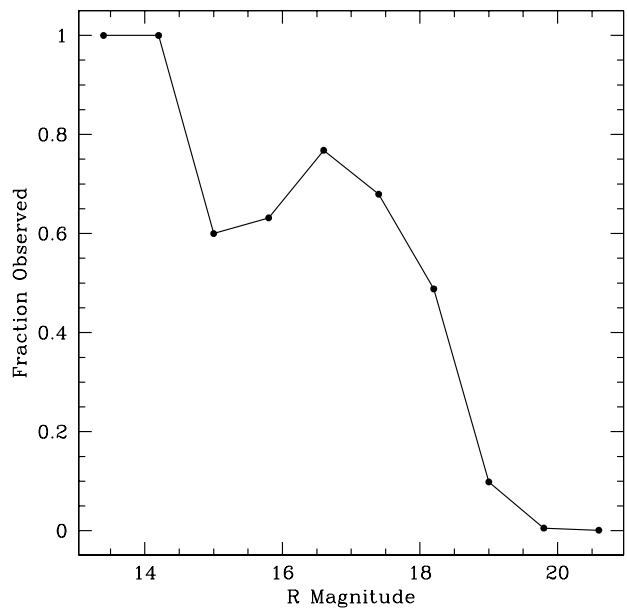,angle=0,width=3.4in}}
  \caption{Fraction of galaxies from the parent APM catalogue
that are selected for observation with 2dF.  Since brighter 
objects are given a higher priority flag in the 2dF configurations,
they are the most likely to have a fibre placed upon them.  
}
\label{fig:select}
\end{figure}

For 2dF observing, each of our target objects is given 
a priority in the {\sc configure} software package
(see www.aao.gov.au/2df/manual/) 
that is proportional to its $R$-band 
magnitude to ensure that the brighter objects (e.g.\ cD type galaxies)
are definitely observed whilst the more common (numerous) fainter
galaxies are less likely to be sampled.  Sky fibres are then allocated
and checked to ensure that they are blank sky and not accidentally 
placed upon actual objects.
Figure~\ref{fig:select} displays the fraction of objects that
are observed out of all possible candidate objects.
At $R<18$, we are at least 60 per cent likely to place a fibre on
a given object.  This fraction drops rapidly toward fainter
magnitudes.

%
%

\subsection{Reduction}

Our data are reduced in
a standard manner in the automated 2dF data reduction pipeline
software (www.aao.gov.au/2df/; see also Bailey et al.\ 2001).

Redshift determination is carried out using the {\sc zcode} package, 
as employed by 2dFGRS (Colless et al.\ 2001) and 2SLAQ 
(Cannon et al.\ 2005).  Briefly, our objects are cross-correlated with
a number of template spectra (including G- and K-type stars; a globular
cluster and several galactic spectra).  Each resultant redshift is
given a Tonry \& Davis (1979) value (TDV) in order to determine which
template spectra provides to best redshift -- usually the one
with the highest TDV.  To check this, each spectra is de-redshifted
to rest frame wavelengths and inspected by eye (KAP \& IGR) to 
check that its emission and absorption features confirm the redshift
determination.  At this stage, we give a 
\emph{qualitative} quality parameter to the
resultant redshift in the range 1 to 4.  A value of 4 denotes a certain
redshift (all spectral features, both emission and
absorption are in the correct place, the spectra has a good
signal to noise ratio and/or TDV with an obvious single cross-correlation 
peak).
A quality of 3 is given to a spectra that is very probably 
correct (it has many or all of the spectral features in the
right locations and a reasonable S/N ratio).  A quality of 2
is given to those spectra that are only 50 per cent likely 
to be correct (typically only two spectral lines are present and
S/N and TDV are low with perhaps two or more plausible cross-correlation
peaks).  A quality of 1 denotes a galaxy that we
could not even guess a redshift for.  We note that all redshifts
used in this work are heliocentric.

\begin{figure}
\centerline{\psfig{file=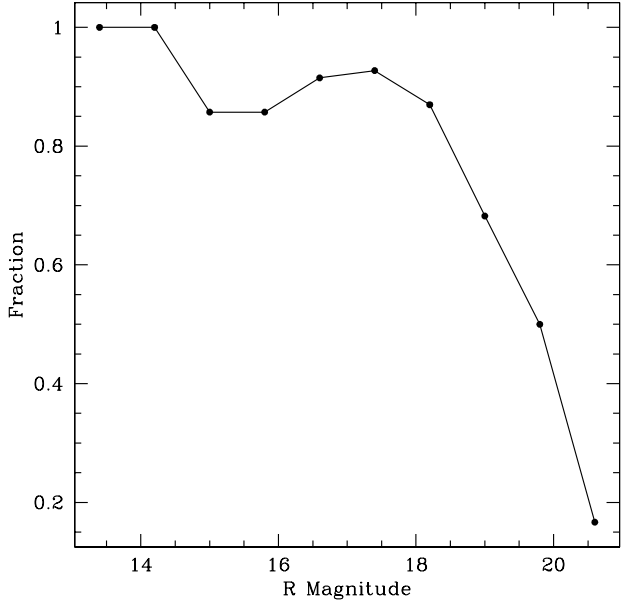,angle=0,width=3.4in}}
  \caption{Spectroscopic success rate, defined as 
the fraction of spectra that
give a quality 3 or 4 redshift, as a function of magnitude.
All of the bright spectra generate a definite redshift,
decreasing to fainter magnitudes.
}
\label{fig:magcomp}
\end{figure}

The fraction of our observations that yield a quality 3 or 4
is shown in Figure~\ref{fig:magcomp} as a function of magnitude.
At bright magnitudes ($R<18$), the redshift completeness is always 
above 80 per cent.  This drops markedly for fainter magnitudes.
Multiplied 
together with Figure~\ref{fig:select}, these fractions define 
our overall survey function (i.e.\ how likely a galaxy is to
be observed and how likely that observation is to produce
a quality redshift).  In principle, there are of course other 
selection and completeness limits in this survey beyond 
Figures~\ref{fig:select} and~\ref{fig:magcomp}.  For example,
only one galaxy in a close pair of galaxies are likely to be 
observed as there is a limit to how close 
2dF can place fibres next to each other without causing problems.
Since we observe with multiple configurations of 2dF, we consider 
such geometric selection effects to be minimized and 
hence consider the product of 
Figures~\ref{fig:select} and~\ref{fig:magcomp}
to be a good approximation for the final overall survey function.  

For the rest of this
work, we will stick to only those galaxies with a quality parameter
of 3 or 4 to ensure that our redshifts are of excellent quality.

%
%

\subsection{Duplicate observations}

Intentionally, observations of some objects are repeated from 
one 2dF configuration to another.  This provides us with a means
to evaluate our own internal error rates.

For our quality 2 objects that have a better quality repeat,
we find that roughly only half of them are within 300kms$^{-1}$ 
of the original measurement.

Out of the 489 objects (of quality 3 or 4) observed, 293 are
unique objects with no secondary observation made.  The remaining
objects are multiple (usually 2 observations, occasionally 3)
observations of a further 98 objects\footnote{Therefore, in total,
we have observed 391 individual objects with a 
quality flag or 3 or 4.  See Appendix~A.}.

For these 98 repeat observations, the median absolute difference in redshift
measurement is $| \Delta z | = 0.00022$ (or about 66 kms$^{-1}$).
Moreover, none of them deviate by more than 300 kms$^{-1}$ from
their duplicate.
This is very comparable to the same measurement performed by
2dFGRS of 64 kms$^{-1}$ (Colless et al.\ 2001) and by LARCS
of 65 kms$^{-1}$ (Pimbblet et al.\ 2006).  The very 
similar values are expected given the highly similar way in which
these surveys all process their datasets.
Significantly, we find no dependence of this value on the redshift
of a given object.  

For our final catalogue (Appendix~A), we remove the duplicated objects.
This is done by selecting the object that has the highest quality flag,
then the highest TDV (in the event of a pair having the same quality)
for inclusion in the final catalogue.  We also 
note that the cD galaxy did not have any repeat observations made.

\section{Redshift Structure}

%
%
\begin{figure}
\centerline{\psfig{file=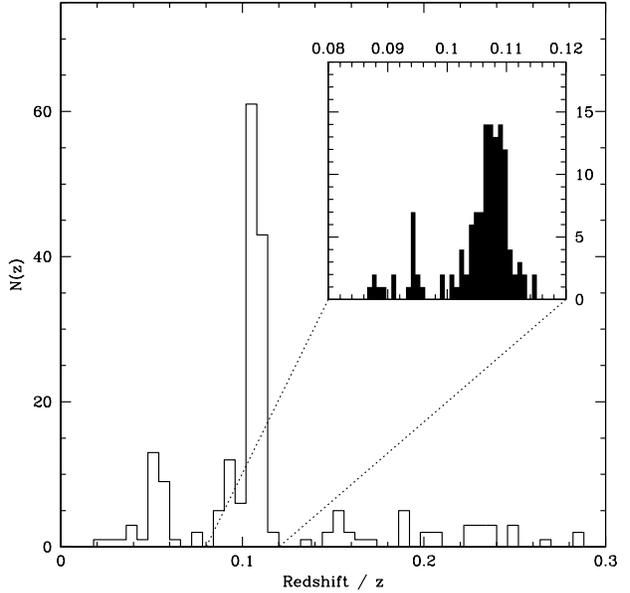,angle=0,width=3.4in}}
  \caption{Redshift histogram for Abell~3653 with an inset
panel depicting a magnification of the central region.
Abell~3653 appears to be a reasonably regular cluster with
a couple of groups in the foreground.  A secondary structure
can also be seen at $z\sim0.05$.  We suggest that this may be part
of Abell~S0835. 
}
\label{fig:zhist}
\end{figure}

%
%
\begin{figure}
\centerline{\psfig{file=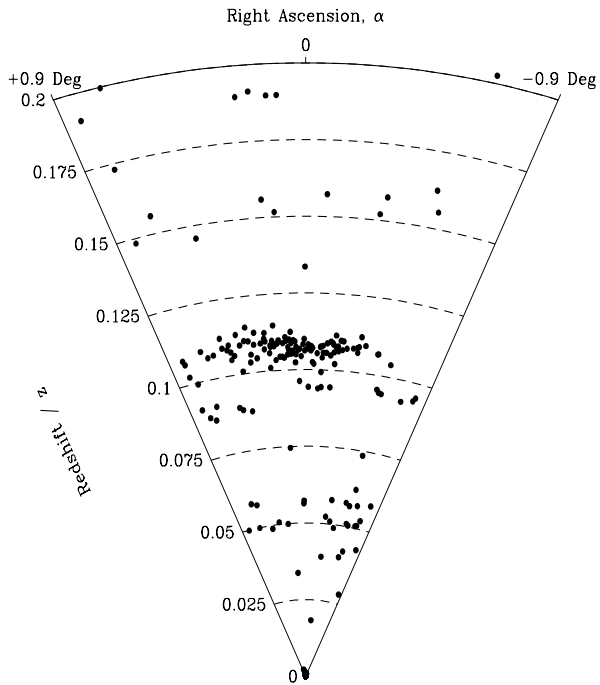,angle=0,width=3.5in,height=4.in}}
\centerline{\psfig{file=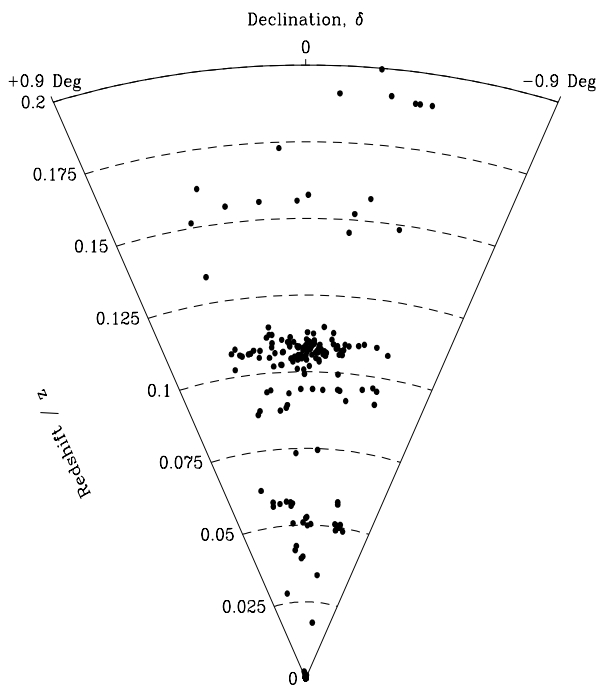,angle=0,width=3.5in,height=4.in}}
  \caption{\small{Wedge plots of right ascension and declination 
versus redshift in the direction of Abell~3653.  
The centre of the
cluster is located at $\alpha = \delta = 0$. 
}}
  \label{fig:wedge}
\end{figure}

Of the 391 objects in our final catalogue, 191 (48.8 per cent)
have redshifts $z<0.01$ and are therefore classified as stars.
For the galaxies in our sample, we now construct a redshift histogram
to qualitatively examine the cluster structure and to aid in 
defining cluster membership (Figure~\ref{fig:zhist}).

Figure~\ref{fig:zhist} suggests that Abell~3653 is a reasonably
regular cluster (approximately a Gaussian shape) 
with only minor sub-structure and a couple of close 
galaxies or groups of galaxies around it in redshift space.  A
secondary structure is also seen at $z\sim0.5$.  
To get a better qualitative assessment of the
overall appearance of these structures, we create wedge plots
of redshift versus the RA and Dec of our observed galaxies
(Figure~\ref{fig:wedge}).

The wedge plots show that the bulk of the cluster is contained
at around $z\approx0.11$, and it appears to be slightly elongated
in RA compared to Dec.  It is also a relatively isolated cluster:
there are almost no galaxies in the immediate background, 
with a couple of galaxies in the foreground.
In the line of sight to the cluster, we again note that there 
are structures at $z\approx0.05$ and $0.06$ which 
do not appear tightly bound
and may therefore be foreground groups 
or perhaps the periphery of Abell~S0835 (located over 
1 degree away from Abell~3653 at $z\sim0.05$).

\subsection{Cluster Membership and Velocity Dispersion}

To define cluster membership we 
employ the statistical clipping technique
of Zabludoff, Huchra \& Geller (1990; ZHG)
to find the cluster's mean velocity, $cz$, and velocity dispersion,
$\sigma_z$.
ZHG is an iterative technique whereby the entire cluster 
population is looped over repeatedly.  Any galaxy
that is more than $1 \sigma_z$ from their nearest neighbour
are flagged for exclusion.  This process is then repeated,
excluding the flagged galaxies, until convergence of the
galaxy population is achieved.

We find that ZHG gives values of $cz = 32214 \pm 83$ kms$^{-1}$ and
$\sigma_{cz} = 880^{+66}_{-54}$, with 111 cluster members.  
The errors on the velocity dispersion and mean
velocity are taken using the formulae presented by
Danese, De Zotti and di Tullio (1980).
This redshift is in good agreement with previous 
estimates (e.g.\ 32670 kms$^{-1}$ from Struble \& Rood 1999)
and confirms that the earlier estimate of Abell, Corwin \& Olowin
(1989) of 14250 kms$^{-1}$ refers to the secondary
structure highlighted by Figure~\ref{fig:zhist} at $z\sim0.05$.

%
%
\begin{figure}
\centerline{\psfig{file=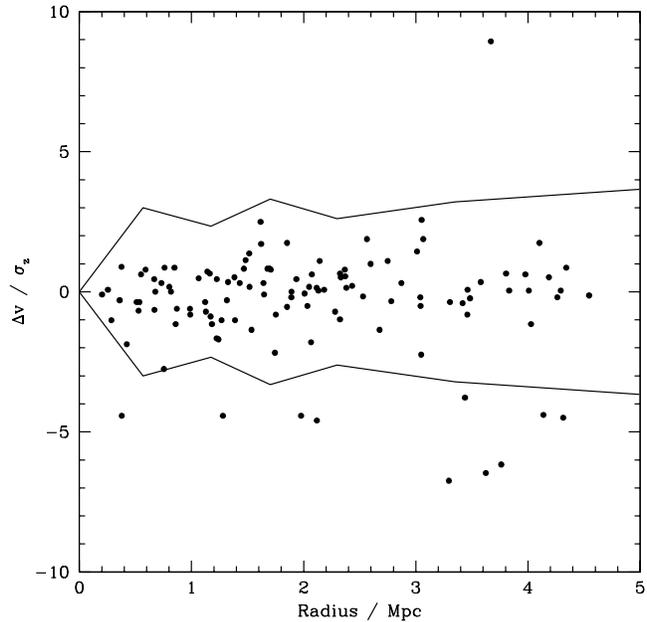,angle=0,width=3.5in}}
  \caption{Phase-space diagram of Abell~3653: radius from the
cluster centre is plotted as a function of velocity difference 
from the cluster mean.  The solid lines are the $3\sigma$ rms
scatter for the cluster members.  All galaxies within these lines
are considered cluster members. 
}
\label{fig:phase}
\end{figure}

\section{cD Peculiar Velocity}

Using the mean velocity of Abell~3653, $cz = 32214 \pm 83$ kms$^{-1}$, 
we now proceed to evaluate the peculiar velocity of the
cluster cD galaxy.  
In Appendix~A, the cD galaxy is recorded as entry PRD196 and has 
a velocity of $cz=32970$ kms$^{-1}$ with an assumed error of
$\pm 66$ kms$^{-1}$ (see above)\footnote{The position of the
cD galaxy is 19 53 3.52  -52 02 16.2, with TDV$=5.4$, a magnitude
of $R=11.2$ and a quality flag of 4.}.
The cD galaxy has also been previously observed by
Postman \& Lauer (1995) who find a velocity of 
$cz=32739 \pm 97$ kms$^{-1}$.  The difference between our
measurement and that of Postman \& Lauer (1995) is 
less that $2\sigma$ and therefore we do not regard it as highly
significant.  Moreover, this reduces the likelihood that the cD 
galaxy possesses multiple nuclei that could contaminate our 
measurement due to random fibre placement.

The errors of the cD galaxy and the cluster are now added 
in quadrature to give a peculiar motion of the cD galaxy of
$756 \pm 106$ kms$^{-1}$ using our velocity for the
cD galaxy.
If we use the velocity
of the cD galaxy found by Postman \& Lauer (1995)
it would be $525 \pm 128$ kms$^{-1}$.  
These values must be corrected by a $(1+z)$ factor
to obtain the peculiar velocity in the cluster rest frame.
This correction gives values of $683 \pm 96$ kms$^{-1}$,
and $474 \pm 116$ kms$^{-1}$, respectively.
We note that these two peculiar 
velocities are well within $2\sigma$ of each other. 
Interestingly, they are \emph{significantly} higher 
than our cluster velocity, by 4.1--7.1$\sigma$.
We therefore reject the assumed null hypothesis 
that the cD galaxy lies at the very centre of our cluster
velocity distribution with a confidence level of $\gg 99.9$ per cent.

Such a cluster restframe peculiar velocity is also rather large
-- we believe that it may be the largest peculiar velocity 
ever reported in the literature.
In Abell~2670, Sharples et al.\ (1988) report a peculiar 
velocity of 439 kms$^{-1}$ -- one of the largest peculiar motions
previously reported in the literature for a cD galaxy.  Indeed, most 
cD galaxies \emph{are} found at the centre of their host cluster
velocity distribution (Oegerle \& Hill 2001; Quintana \& 
Lawrie 1982).  
We note, however, that these studies have used cluster galaxies
that have a much more confined radial extent from the cluster 
centre (or a different magnitude limit)
than those presented in this work, generally at least
within 3 Mpc.
To ascertain if this has any effect on the cD peculiar velocity,
we now re-compute the cluster's velocity using the ZHG method,
but restrict our sample in both radius and
magnitude.  The results of this
analysis are presented in Table~\ref{tab:zhg}.  Neither 
restricting the sample in radius from the cluster centre
or magnitude makes the peculiar velocity of the cD galaxy 
insignificant.

%
%
\begin{table*}
\begin{center}
\caption{Cluster velocity measurements via ZHG for 
a variety of samples.  The column headed `cD $cz$ significance' 
lists how significant our measurement of the peculiar velocity of
the cD galaxy is away from the cluster velocity measurement for each
given sample. 
\hfil}
\begin{tabular}{lcccc}
\noalign{\medskip}
\hline
Sample      & N(gal) & cz           & $\sigma_{cz}$  & cD $cz$ \\
Restriction &        & (kms$^{-1}$) & (kms$^{-1}$) &  significance \\
\hline
unrestricted & 111   & $32214 \pm 83$  & $880^{+66}_{-54}$  & $7.1\sigma$ \\
\hline
$r<1.0$ Mpc & 24     & $31986 \pm 159$ & $780^{+147}_{-94}$ & $5.7\sigma$ \\
$r<2.0$ Mpc & 59     & $32154 \pm 114$ & $878^{+95}_{-71}$  & $6.2\sigma$ \\
$r<3.0$ Mpc & 83     & $32207 \pm 92$  & $837^{+74}_{-58}$  & $6.7\sigma$ \\
\hline
$R<17.5$    & 33     & $32155 \pm 147$ & $844^{+129}_{-88}$ & $5.1\sigma$ \\
$R<18.0$    & 60     & $32152 \pm 110$ & $855^{+91}_{-69}$  & $6.4\sigma$ \\
$R<18.5$    & 90     & $32214 \pm 87$  & $827^{+70}_{-56}$  & $6.9\sigma$ \\
\hline
\noalign{\smallskip}
\end{tabular}
  \label{tab:zhg}
\end{center}
\end{table*}

\section{Cluster Morphology}

%
%
\begin{figure}
\centerline{\psfig{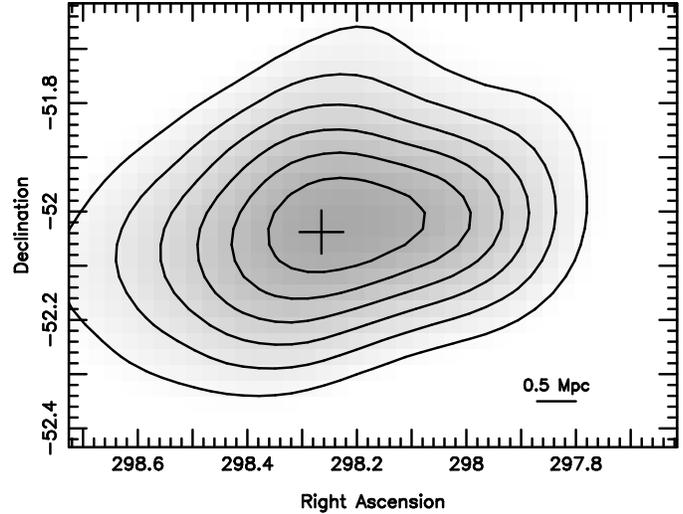}}
  \caption{A smoothed spatial distribution of galaxies
belonging to Abell~3653.  The positions of the galaxies are
smoothed with a Gaussian kernal of length 0.5 Mpc at the 
mean redshift of the cluster (scale bar in the lower right).  
The lowest contour represents
a surface density of 3 Mpc$^{-2}$ and each contour interior
to this increases by 1 Mpc$^{-2}$. 
The central crosshair denotes the location of the cD galaxy.
The cluster displays
a mild elongation in RA compared to Dec., but is otherwise
quite regular morphologically. 
}
\label{fig:map}
\end{figure}

In Figure~\ref{fig:map} we plot the positions of cluster members
smoothed with a Gaussian kernel of length 0.5 Mpc.  The cluster
appears to be quite regular in shape, with only a mild elongation
in right ascension compared to declination.
We also note that the cD galaxy is located just off 
the NED adopted cluster centre by about 0.7 Mpc.

To better probe the internal structure of the cluster,
we now apply the Dressler \& Shectman (1988; DS herein)
test to search for any sign of substructure.
We elect to use the DS approach as a 
lack of consensus concerning the optimal
tool to use for the detection of substructure led Pinkney et al.\ (1996)
to investigate differences among a battery of statistical tools.  They
conclude by endorsing the DS test as by far 
the most powerful and most sensitive (three-dimensional)
substructure detection tool available in the general case.
The DS method is briefly outlined below.

For each cluster member, its ten nearest
neighbours on the sky
are found.  For this group of 11 galaxies, the \emph{local}
mean velocity, $\overline{v}_{local}$, and velocity dispersion, 
$\sigma_{local}$ is calculated and compared to the corresponding global 
values.  Thus the deviation is:

\begin{equation}
\delta^2_i = (11/\sigma^2) [ (\overline{v}_{local} - \overline{v})^2 + (\sigma_{local} - \sigma)^2 ]
\end{equation}

The parameter of merit, however, is the cumulative deviation, $\Delta$,
which is the sum of all $\delta_i$ for the $N$ cluster members:

\begin{equation}
\Delta = \sum_{i=1}^{N} \delta_i
\end{equation}

For a cluster velocity dispersion which is Gaussian in nature, the
$\Delta$ statistic will be of order $N$.  
As the underlying distribution may not be Gaussian even for a 
cluster without substructure, 1000 Monte-Carlo simulation are run on 
the clusters, each time shuffling the galaxies velocities but holding
their positions constant.
Caution must be exercised as should substructure be 
aligned along the line of sight to the cluster, the $\Delta$ statistic 
will not show this.

\begin{figure*}
\centerline{\psfig{file=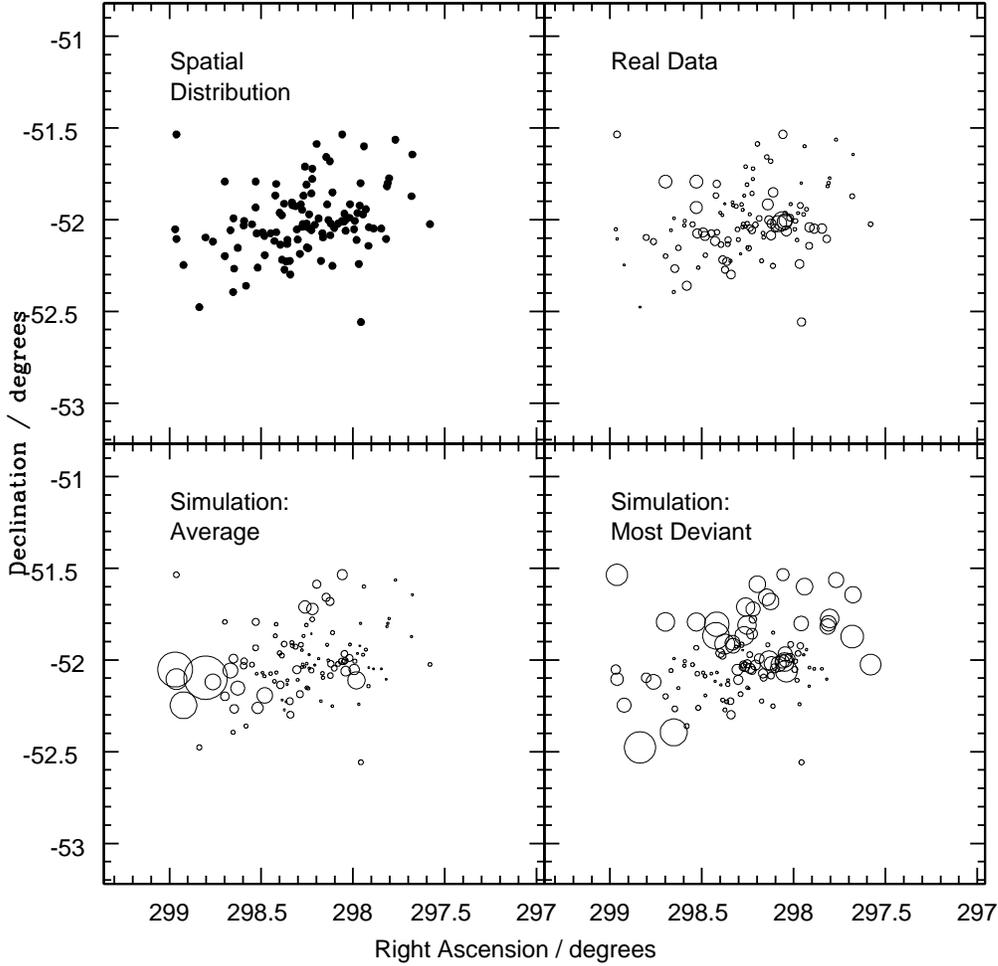,angle=0,width=5.5in}}
  \caption{Spatial distribution 
and the results of the DS tests.
Top Left: the spatial distribution of cluster members. 
Top Right: DS test of the actual data.  
Bottom Left: average of the 1000 Monte-Carlo simulations.  
Bottom Right: most deviant of the 1000 Monte-Carlo simulations.
Abell~3653 does not exhibit any significant subclustering.  
}
\label{fig:dstest}
\end{figure*}

From the DS test we obtain $N=111$ and find
that $\Delta_{Obs} = 135$, with  
the average simulation producing $\Delta_{Av} = 122$,
and the most deviant simulation giving $\Delta_{max} = 178$.
This yields $P(\Delta) = 0.169$ meaning that Abell~3653 does
not have any significant substructure over the entire spatial range
probed.  Moreover, $P(\Delta)$ does not become significant
(i.e.\ $<0.001$) over any range of radius from the cluster centre.
Our result from the DS test is
illustrated in 
Figure~\ref{fig:dstest}.
Each galaxy is marked with a circle whose diameter is proportional 
to $e^{\delta}$, therefore any subclustering would 
be seen as a localized group of overlapping circles.
Figure~\ref{fig:dstest} confirms that there is no significant 
substructuring within the cluster.

%
%

\section{Discussion}

It is certain that the presence of substructure will affect
the cluster velocity and velocity dispersion -- thereby perturbing
the peculiar velocity measurement of the cD galaxy.  
Equally, during a cluster-cluster merger, the cD galaxy may
be relocated from its assumed position at the bottom of the
gravitational potential well of the cluster (Zabludoff \& Zaritsky 1995). 
However, since Abell~3653 shows no sign of significant subclustering
on any scales (i.e.\ in any radially-limited subset of 
the cluster members), we conclude that the peculiar velocity of
the cD galaxy cannot be due to perturbations of the mean
cluster velocity arising from infalling galaxy groups or subclusters. 

It is possible that the cD galaxy is a part of a very localized subclump.
There are 9 galaxies in close proximity (within 5 arcmin) 
of the cD galaxy.  Using ZHG, the systemic velocity of these 9 galaxies
is $32323 \pm 204$ kms$^{-1}$.  Thus the cluster restframe 
peculiar velocity of the cD galaxy to those galaxies in close
proximity to it is $584 \pm 214$ kms$^{-1}$.   
This is just under $3\sigma$ and gives a confidence level of
$>99.5$ per cent to reject the null hypothesis that the cD galaxy is
at the centre of this local group of galaxies. 
Although the DS test has been endorsed as one of the best
tests around for finding substructure (Pinkney et al.\ 1996),
we can consider if it would be capable of detecting a small
group around the cD galaxy.
We do this by giving the 9 galaxies around the cD galaxy a velocity 
appropriate to a small group (say up to $\sigma_{cz}$=350kms$^{-1}$),
repeating the DS test and then 
performing a Monte-Carlo simulation of this analysis 100 times.
We find that the DS test is able to pick them up as a significant
substructure within the cluster in all cases.

Within this sample, however, there are 3 galaxies that have
redshifts that are very close to that of the cD galaxy.  
It may be the case that the cD galaxy formed very early in the
evolution of a cluster (Merritt 1985).
Consider a small group that collapses and subsequently virializes.
Such a system may later encounter other similar systems and merge 
with them to form a richer cluster 
(Sharples et al.\ 1988; White \& Rees 1978).
Since the cD galaxy has a large peculiar velocity, it may
simply be the case that it has not had sufficient time
to complete the mixing process; akin to a more advanced 
stage of Abell~2670 (Sharples et al.\ 1988).
Our results favour a picture of cluster growth fuelled by
the hierarchical accretion of sub-clusters and galaxy groups.
Conversely, they strongly argue against a picture of 
cD galaxies that grow in situ, post-virialization of the cluster.

\section{Summary}

This work presents new observations of the galaxy cluster Abell~3653
using the 2dF spectrograph to amass 391 new redshift measurements.
Of these, we find 111 are bone-fide cluster members, whilst about 
half of them are stars oweing to our non-biased approach to 
selecting objects for observation.
Our main findings are:

\begin{itemize}

\item Abell~3653 has a velocity of $cz = 32214 \pm 83$ kms$^{-1}$
($z=0.10738 \pm 0.00027$),
with a velocity dispersion of $\sigma_{cz} = 880^{+66}_{-54}$. 
Such a value of $\sigma_{cz}$ is typical for massive clusters
at these redshifts (cf.\ Pimbblet et al.\ 2006).

\item From our data, the cD galaxy of Abell~3653 has an 
extremely large cluster restframe 
peculiar velocity of $683 \pm 96$ kms$^{-1}$.
This is over $7\sigma$ away from the mean cluster velocity and
makes the peculiar velocity of this cD galaxy 
one of the largest, perhaps even the 
largest ever recorded for a cD galaxy. 
The significance and magnitude of the peculiar velocity 
do not change significantly by restricting our sample to
brighter galaxies and / or smaller radii from the cluster
centre.

\item Using a DS test, 
Abell~3653 shows no sign of significant subclustering 
on any length scale.  Therefore the peculiar velocity of
the cD galaxy cannot be accounted for by considering recent
(major) cluster merger events.

\item Our results favour a scenario whereby rich clusters
grow in a hierarchical fashion from sub-clumps and groups
infalling into a gravitational potential well.  
They do not favour theories which require a cD galaxy
to sit at the bottom of a gravitational 
potential well and grow in situ there.

\end{itemize}

\section*{Acknowledgments}

We warmly thank the 2dF observers who carried out the observations
for us in service mode at the AAT, namely Rob Sharp and Quentin
Parker.
We also wish to thank Michael Drinkwater for stimulating
conversations that have improved this work and an anonymous
referee for very useful feedback.
KAP acknowledges support from an EPSA University of Queensland Research
Fellowship.

\section*{Appendix~A: The Redshift Catalogue}

Table~\ref{tab:cat} gives the final redshift catalogue for
all objects observed in the direction of Abell~3653 with a 
quality flag of 3 or 4. 
Note that all redshifts are heliocentric. 
The full version of the spectroscopic
catalogue will also be made available in Synergy, the online version
of the Monthly Notices of the Royal Astronomical Society.

\newpage
%
%
\begin{table*}
\begin{center}
\caption{The redshift catalogue.  The R magnitude is sourced 
from the APM catalogue.  Rather than providing an error on
the redshift, z, which would only be the error on the cross-correlation
fit, we follow 2dFGRS data release model to 
give the Tonry-Davis value (TDV) and our own 
quality assessment parameter.  
\hfil}
\begin{tabular}{lllccccl}
\noalign{\medskip}
\hline
Identification & RA & Dec & R & z & TDV & Quality & Abell~3653 \\
Tag            & (J2000) & (J2000) & &  & &      & Member? \\
\hline
PRD1 & 19 55 15.27 & -51 58 7.6 & 18.66 & 0.2279 & 6.04 & 4 \\
PRD2 & 19 55 10.84 & -51 53 25.5 & 18.12 & 0.2848 & 9.30 & 4 \\
PRD3 & 19 54 7.32 & -51 56 1.8 & 18.22 & 0.1034 & 6.22 & 4 & yes \\
PRD4 & 19 55 12.57 & -51 50 16.1 & 18.96 & 0.3220 & 6.91 & 4 \\
PRD5 & 19 55 59.11 & -51 37 41.4 & 18.16 & 0.2358 & 5.47 & 3 \\
PRD6 & 19 53 15.11 & -51 55 38.8 & 18.78 & 0.1044 & 6.37 & 3 & yes \\
PRD7 & 19 53 30.05 & -51 54 48.2 & 18.37 & 0.1098 & 6.26 & 3 & yes \\
PRD8 & 19 54 12.13 & -51 48 29.3 & 18.85 & 0.0003 & 3.76 & 3 \\
PRD9 & 19 54 57.22 & -51 34 14.8 & 18.36 & -0.0002 & 5.61 & 3 \\
PRD10 & 19 52 45.03 & -51 59 35.6 & 17.20 & 0.1044 & 7.20 & 3 & yes \\
PRD11 & 19 53 25.81 & -51 48 36.1 & 18.39 & 0.1559 & 6.82 & 3 \\
PRD12 & 19 54 9.12 & -51 35 26.8 & 18.14 & -0.0004 & 4.54 & 3 \\
PRD13 & 19 54 3.77 & -52 00 29.3 & 18.45 & -0.0003 & 5.14 & 3 \\
PRD14 & 19 52 26.82 & -51 51 5.0 & 17.36 & 0.1025 & 6.25 & 3 & yes \\
PRD15 & 19 52 40.87 & -51 54 14.0 & 18.80 & 0.0004 & 9.01 & 4 \\
PRD16 & 19 52 18.02 & -51 30 41.5 & 17.93 & 0.0001 & 4.78 & 3 \\
PRD17 & 19 52 14.75 & -51 39 51.0 & 18.28 & 0.0001 & 4.39 & 4 \\
PRD18 & 19 52 32.37 & -52 00 8.1 & 18.14 & 0.1071 & 6.90 & 3 & yes \\
PRD19 & 19 50 55.29 & -51 41 12.6 & 18.10 & 0.0001 & 5.43 & 3 \\
PRD20 & 19 51 8.84 & -51 32 42.7 & 17.65 & 0.0002 & 12.80 & 4 \\
PRD21 & 19 51 16.34 & -51 29 49.4 & 17.90 & -0.0003 & 3.97 & 3 \\
PRD22 & 19 50 9.19 & -51 33 57.8 & 17.22 & 0.0002 & 5.93 & 3 \\
PRD23 & 19 51 12.93 & -51 46 26.7 & 18.53 & 0.1116 & 3.03 & 3 & yes \\
PRD24 & 19 51 14.49 & -51 48 1.2 & 18.67 & 0.1083 & 7.18 & 3 & yes \\
PRD25 & 19 49 44.42 & -51 53 40.4 & 17.48 & 0.0931 & 3.30 & 4 \\
PRD26 & 19 51 43.20 & -51 56 37.5 & 17.61 & 0.1098 & 6.79 & 3 & yes \\
PRD27 & 19 49 31.34 & -51 40 59.2 & 18.37 & 0.0000 & 5.13 & 3 \\
PRD28 & 19 50 10.10 & -51 45 49.7 & 18.14 & 0.0001 & 4.12 & 3 \\
PRD29 & 19 52 10.47 & -52 00 14.8 & 17.94 & 0.1079 & 7.78 & 3 & yes \\
PRD30 & 19 49 39.52 & -52 02 32.2 & 18.49 & 0.0003 & 13.70 & 4 \\
PRD31 & 19 51 23.09 & -52 02 54.6 & 17.92 & 0.1076 & 7.05 & 3 & yes \\
PRD32 & 19 50 27.53 & -52 15 27.8 & 17.38 & 0.0945 & 5.24 & 3 \\
PRD33 & 19 50 49.45 & -52 09 5.6 & 18.45 & 0.2504 & 7.60 & 4 \\
PRD34 & 19 49 34.02 & -52 23 21.7 & 16.84 & -0.0002 & 5.24 & 3 \\
PRD35 & 19 52 24.52 & -52 02 43.7 & 18.78 & 0.1019 & 6.80 & 3 & yes \\
PRD36 & 19 50 15.31 & -52 23 17.0 & 18.19 & 0.0000 & 5.21 & 3 \\
PRD37 & 19 50 23.25 & -52 29 51.1 & 17.72 & 0.0509 & 3.31 & 3 \\
PRD38 & 19 50 35.29 & -52 29 0.2 & 18.28 & 0.0000 & 4.41 & 4 \\
PRD39 & 19 50 53.30 & -52 25 35.8 & 18.31 & 0.0002 & 4.99 & 3 \\
PRD40 & 19 50 59.64 & -52 26 17.7 & 18.25 & -0.0002 & 5.94 & 3 \\
PRD41 & 19 51 45.13 & -52 25 38.4 & 18.40 & 0.0002 & 8.02 & 4 \\
PRD42 & 19 51 6.75 & -52 25 12.7 & 19.17 & 0.2048 & 6.23 & 3 \\
PRD43 & 19 51 35.87 & -52 32 4.5 & 18.70 & 0.2314 & 7.43 & 4 \\
PRD44 & 19 51 55.55 & -52 06 40.8 & 18.04 & 0.1089 & 5.80 & 3 & yes \\
PRD45 & 19 52 40.95 & -52 24 12.5 & 17.60 & 0.0000 & 6.53 & 3 \\
PRD46 & 19 52 48.57 & -52 30 38.0 & 18.78 & 0.0963 & 3.83 & 3 \\
PRD47 & 19 52 50.02 & -52 27 26.5 & 18.10 & 0.0002 & 5.55 & 3 \\
PRD48 & 19 52 39.28 & -52 05 49.2 & 17.70 & 0.1063 & 4.29 & 4 & yes \\
PRD49 & 19 52 58.66 & -52 09 19.7 & 18.32 & 0.1063 & 6.78 & 3 & yes \\
PRD50 & 19 52 40.06 & -52 04 24.7 & 18.23 & 0.1100 & 6.80 & 3 & yes \\
PRD51 & 19 53 21.90 & -52 17 56.3 & 17.83 & 0.1089 & 7.16 & 3 & yes \\
PRD52 & 19 54 36.77 & -52 23 41.2 & 18.05 & 0.1099 & 4.00 & 3 & yes \\
PRD53 & 19 54 51.28 & -52 22 3.9 & 17.55 & 0.0001 & 5.68 & 3 \\
PRD54 & 19 53 12.18 & -52 07 55.2 & 16.02 & 0.0746 & 4.79 & 3 \\
PRD55 & 19 55 20.96 & -52 28 37.2 & 17.07 & 0.1095 & 6.76 & 3 & yes \\
PRD56 & 19 55 6.36 & -52 30 29.5 & 17.52 & 0.0003 & 6.26 & 3 \\
PRD57 & 19 53 25.84 & -52 06 38.8 & 17.90 & 0.1034 & 6.34 & 3 & yes \\
PRD58 & 19 54 18.48 & -52 11 35.6 & 18.11 & 0.2066 & 4.80 & 3 \\
PRD59 & 19 55 51.10 & -52 19 40.7 & 17.72 & 0.0911 & 3.43 & 3 \\
PRD60 & 19 54 16.35 & -52 10 38.2 & 18.49 & 0.0001 & 24.50 & 4 \\
\hline
\noalign{\smallskip}
\end{tabular}
\end{center}
\end{table*}

%
%
\begin{table*}
\setcounter{table}{2}
\begin{center}
\caption{continued.
\hfil}
\begin{tabular}{lllccccl}
\noalign{\medskip}
\hline
Identification & RA & Dec & R & z & TDV & Quality & Abell~3653 \\
Tag            & (J2000) & (J2000) & &  & &      & Member? \\
\hline
PRD61 & 19 53 4.00 & -52 04 23.3 & 17.59 & 0.0001 & 4.69 & 3 \\
PRD62 & 19 55 41.49 & -52 09 31.5 & 18.65 & 0.0001 & 4.78 & 3 \\
PRD63 & 19 55 57.85 & -52 09 45.7 & 18.94 & 0.2862 & 5.23 & 4 \\
PRD64 & 19 55 12.77 & -52 05 50.0 & 17.93 & 0.1070 & 6.40 & 3 & yes \\
PRD65 & 19 55 58.62 & -52 08 45.7 & 18.30 & 0.1910 & 8.39 & 3 \\
PRD66 & 19 54 5.99 & -52 04 31.2 & 17.78 & 0.1103 & 7.78 & 3 & yes \\
PRD67 & 19 55 50.86 & -52 06 18.5 & 19.48 & 0.1065 & 3.48 & 3 & yes \\
PRD68 & 19 54 7.45 & -52 00 23.5 & 17.16 & 0.0001 & 5.38 & 4 \\
PRD69 & 19 55 45.79 & -51 54 43.2 & 18.43 & 0.1731 & 4.31 & 3 \\
PRD70 & 19 54 42.29 & -51 52 37.4 & 18.05 & 0.0893 & 6.59 & 3 \\
PRD71 & 19 55 60.00 & -51 43 54.8 & 18.21 & -0.0003 & 4.48 & 3 \\
PRD72 & 19 54 47.58 & -51 47 33.7 & 20.56 & 0.1125 & 2.76 & 3 & yes \\
PRD73 & 19 55 36.62 & -51 42 2.2 & 18.34 & 0.0002 & 5.96 & 4 \\
PRD74 & 19 56 2.33 & -51 38 28.9 & 18.09 & 0.0000 & 10.81 & 4 \\
PRD75 & 19 54 26.77 & -51 45 46.6 & 18.14 & 0.0002 & 9.85 & 4 \\
PRD76 & 19 55 40.02 & -51 31 47.7 & 17.38 & 0.0001 & 14.54 & 4 \\
PRD77 & 19 53 19.94 & -51 54 21.1 & 17.09 & 0.1098 & 7.51 & 3 & yes \\
PRD78 & 19 53 8.08 & -51 55 2.3 & 18.01 & 0.1095 & 7.54 & 3 & yes \\
PRD79 & 19 53 41.68 & -51 52 6.9 & 18.22 & 0.1106 & 4.89 & 4 & yes \\
PRD80 & 19 53 40.52 & -51 48 19.3 & 17.41 & 0.1097 & 7.26 & 3 & yes \\
PRD81 & 19 54 41.11 & -51 35 7.4 & 17.56 & 0.0000 & 4.39 & 4 \\
PRD82 & 19 52 53.13 & -51 43 20.2 & 18.09 & 0.1075 & 7.52 & 3 & yes \\
PRD83 & 19 52 47.47 & -51 35 14.9 & 18.07 & 0.1068 & 7.75 & 3 & yes \\
PRD84 & 19 52 42.56 & -51 42 21.6 & 16.95 & 0.0573 & 5.02 & 4 \\
PRD85 & 19 52 16.52 & -51 57 17.7 & 18.28 & 0.0000 & 7.31 & 3 \\
PRD86 & 19 52 10.63 & -51 45 40.4 & 17.53 & 0.0944 & 4.21 & 4 \\
PRD87 & 19 51 45.78 & -51 36 1.0 & 18.01 & 0.1063 & 2.85 & 4 & yes \\
PRD88 & 19 51 37.69 & -51 36 57.7 & 18.99 & 0.2322 & 8.03 & 4 \\
PRD89 & 19 51 28.41 & -51 46 21.9 & 18.59 & 0.0000 & 4.09 & 3 \\
PRD90 & 19 52 4.00 & -51 54 57.0 & 17.49 & 0.1087 & 7.40 & 3 & yes \\
PRD91 & 19 50 22.37 & -51 41 30.3 & 18.02 & 0.0008 & 5.03 & 3 \\
PRD92 & 19 50 13.20 & -51 39 22.1 & 18.30 & 0.1554 & 7.27 & 3 \\
PRD93 & 19 51 51.32 & -51 55 22.6 & 18.15 & 0.1114 & 7.21 & 4 & yes \\
PRD94 & 19 49 20.70 & -51 38 18.7 & 18.30 & 0.0004 & 4.46 & 4 \\
PRD95 & 19 50 39.07 & -51 50 9.9 & 18.25 & 0.0576 & 3.82 & 4 \\
PRD96 & 19 50 8.81 & -51 50 46.1 & 18.14 & 0.0419 & 3.52 & 4 \\
PRD97 & 19 50 4.47 & -51 54 13.9 & 18.24 & 0.0000 & 5.34 & 4 \\
PRD98 & 19 51 54.69 & -51 59 1.5 & 18.26 & 0.0944 & 6.65 & 3 \\
PRD99 & 19 50 18.56 & -51 58 12.7 & 19.32 & 0.0398 & 3.70 & 4 \\
PRD100 & 19 49 37.91 & -51 57 52.5 & 17.40 & 0.0003 & 4.55 & 4  \\
PRD101 & 19 52 29.00 & -52 01 17.0 & 17.77 & 0.1076 & 7.02 & 3 & yes \\
PRD102 & 19 51 34.14 & -52 01 20.4 & 18.07 & 0.0523 & 3.64 & 3 \\
PRD103 & 19 50 8.13 & -52 08 48.5 & 18.29 & -0.0003 & 4.59 & 3 \\
PRD104 & 19 51 26.98 & -52 04 47.6 & 17.84 & -0.0001 & 14.56 & 4 \\
PRD105 & 19 50 16.76 & -52 05 22.8 & 16.45 & 0.0504 & 5.03 & 4 \\
PRD106 & 19 49 33.98 & -52 05 51.4 & 18.47 & 0.0000 & 11.86 & 4 \\
PRD107 & 19 49 14.51 & -52 06 14.0 & 18.23 & 0.0001 & 4.89 & 3 \\
PRD108 & 19 49 57.38 & -52 13 38.1 & 18.18 & 0.2026 & 4.41 & 3 \\
PRD109 & 19 49 20.37 & -52 18 35.4 & 17.93 & 0.0002 & 5.42 & 3 \\
PRD110 & 19 51 49.02 & -52 03 59.9 & 18.40 & 0.0000 & 4.61 & 4 \\
PRD111 & 19 50 22.49 & -52 16 21.7 & 18.13 & 0.0942 & 5.57 & 4 \\
PRD112 & 19 50 53.09 & -52 14 18.0 & 18.17 & 0.0003 & 4.13 & 3 \\
PRD113 & 19 51 15.39 & -52 14 46.5 & 18.42 & 0.1520 & 7.67 & 3 \\
PRD114 & 19 50 11.94 & -52 26 46.1 & 18.23 & 0.0002 & 4.92 & 4 \\
PRD115 & 19 50 57.12 & -52 19 28.6 & 18.41 & -0.0001 & 4.47 & 4 \\
PRD116 & 19 51 16.92 & -52 16 9.2 & 17.38 & 0.2497 & 11.96 & 4 \\
PRD117 & 19 51 50.94 & -52 16 13.2 & 19.38 & 0.0183 & 5.19 & 4 \\
PRD118 & 19 51 21.16 & -52 29 37.9 & 18.23 & 0.0003 & 4.28 & 3 \\
PRD119 & 19 51 40.41 & -52 14 11.0 & 18.11 & -0.0002 & 4.92 & 3 \\
PRD120 & 19 52 27.11 & -52 15 7.3 & 17.94 & 0.1071 & 7.88 & 3 & yes \\
PRD121 & 19 52 41.98 & -52 13 28.2 & 18.89 & 0.1083 & 6.45 & 3 & yes \\
PRD122 & 19 53 13.25 & -52 26 38.9 & 19.01 & 0.1897 & 5.17 & 3 \\
PRD123 & 19 53 15.76 & -52 15 22.0 & 18.06 & 0.0338 & 3.14 & 4 \\
\hline
\noalign{\smallskip}
\end{tabular}
\end{center}
\end{table*}

%
%
\begin{table*}
\setcounter{table}{2}
\begin{center}
\caption{continued.
\hfil}
\begin{tabular}{lllccccl}
\noalign{\medskip}
\hline
Identification & RA & Dec & R & z & TDV & Quality & Abell~3653 \\
Tag            & (J2000) & (J2000) & &  & &      & Member? \\
\hline
PRD124 & 19 53 32.76 & -52 13 6.8 & 18.56 & 0.1078 & 6.29 & 3 & yes \\
PRD125 & 19 55 50.32 & -52 29 13.9 & 18.79 & 0.2646 & 6.72 & 3 \\
PRD126 & 19 56 0.70 & -52 29 14.3 & 17.72 & 0.0003 & 5.40 & 3 \\
PRD127 & 19 53 55.42 & -52 11 37.1 & 16.98 & 0.1129 & 5.31 & 4 & yes \\
PRD128 & 19 55 52.87 & -52 27 49.8 & 18.26 & 0.1484 & 5.52 & 3 \\
PRD129 & 19 55 26.34 & -52 21 35.6 & 18.14 & 0.0003 & 13.43 & 4 \\
PRD130 & 19 53 35.00 & -52 08 10.1 & 17.15 & 0.1058 & 7.47 & 3 & yes \\
PRD131 & 19 55 45.95 & -52 13 0.1 & 18.14 & -0.0005 & 5.01 & 3 \\
PRD132 & 19 53 13.44 & -52 03 12.4 & 18.27 & 0.1088 & 3.17 & 4 & yes \\
PRD133 & 19 53 5.80 & -52 01 30.0 & 18.41 & 0.1074 & 7.65 & 3 & yes \\
PRD134 & 19 55 39.81 & -52 05 27.1 & 18.44 & -0.0001 & 5.56 & 3 \\
PRD135 & 19 54 40.34 & -52 03 28.4 & 18.04 & 0.1084 & 6.27 & 3 & yes \\
PRD136 & 19 53 52.80 & -52 02 14.4 & 18.24 & 0.0003 & 13.47 & 4 \\
PRD137 & 19 54 45.19 & -52 02 2.7 & 18.20 & 0.0002 & 5.62 & 3 \\
PRD138 & 19 54 12.17 & -52 01 33.0 & 17.66 & 0.1106 & 5.89 & 3 & yes \\
PRD139 & 19 54 54.94 & -52 02 6.7 & 17.19 & 0.0000 & 6.05 & 3 \\
PRD140 & 19 54 23.05 & -52 01 53.9 & 18.15 & 0.1129 & 4.36 & 3 & yes \\
PRD141 & 19 54 36.38 & -51 59 34.4 & 18.24 & 0.1050 & 5.01 & 3 & yes \\
PRD142 & 19 54 41.86 & -51 43 38.3 & 18.68 & 0.2388 & 4.13 & 3 \\
PRD143 & 19 54 36.62 & -51 51 59.1 & 19.85 & 0.0884 & 3.35 & 4 \\
PRD144 & 19 53 19.20 & -51 55 20.5 & 16.42 & 0.1044 & 6.24 & 3 & yes \\
PRD145 & 19 55 40.98 & -51 36 15.1 & 17.52 & 0.0002 & 4.57 & 3 \\
PRD146 & 19 55 55.23 & -51 32 40.4 & 16.36 & -0.0002 & 5.34 & 3 \\
PRD147 & 19 55 51.01 & -51 32 9.4 & 19.31 & 0.1023 & 3.00 & 4 & yes \\
PRD148 & 19 53 18.89 & -51 51 27.5 & 13.69 & 0.0000 & 5.98 & 3 \\
PRD149 & 19 54 26.86 & -51 34 57.9 & 18.45 & 0.2335 & 5.00 & 3 \\
PRD150 & 19 54 36.58 & -51 29 40.0 & 17.88 & 0.0000 & 5.73 & 3 \\
PRD151 & 19 53 6.73 & -51 56 48.4 & 17.12 & 0.1050 & 5.08 & 4 & yes \\
PRD152 & 19 53 34.75 & -51 39 49.9 & 17.63 & 0.0003 & 5.82 & 3 \\
PRD153 & 19 52 53.34 & -51 46 42.8 & 19.42 & 0.1050 & 3.73 & 4 & yes \\
PRD154 & 19 52 30.20 & -51 40 52.7 & 19.74 & 0.1078 & 4.11 & 4 & yes \\
PRD155 & 19 52 18.71 & -51 29 23.8 & 17.79 & -0.0003 & 4.14 & 3 \\
PRD156 & 19 52 7.18 & -51 33 11.4 & 18.70 & -0.0007 & 4.09 & 3 \\
PRD157 & 19 51 0.35 & -51 30 26.4 & 13.76 & 0.0001 & 4.31 & 3 \\
PRD158 & 19 50 21.23 & -51 30 59.0 & 18.84 & 0.0623 & 3.22 & 3 \\
PRD159 & 19 51 24.06 & -51 30 30.0 & 17.38 & 0.0002 & 4.40 & 4 \\
PRD160 & 19 50 20.41 & -51 32 53.4 & 18.53 & 0.1623 & 6.35 & 3 \\
PRD161 & 19 50 37.27 & -51 38 4.0 & 18.86 & 0.2371 & 3.47 & 4 \\
PRD162 & 19 49 49.16 & -51 29 20.5 & 16.11 & 0.0004 & 8.56 & 4 \\
PRD163 & 19 49 47.32 & -51 30 33.4 & 16.31 & 0.0002 & 5.11 & 4 \\
PRD164 & 19 51 47.99 & -51 37 3.1 & 19.12 & 0.2279 & 3.14 & 4 \\
PRD165 & 19 50 25.06 & -51 41 6.2 & 13.99 & -0.0004 & 4.28 & 3 \\
PRD166 & 19 50 2.20 & -51 36 5.2 & 16.46 & 0.0002 & 5.53 & 4 \\
PRD167 & 19 50 30.41 & -51 50 31.9 & 17.26 & 0.0002 & 4.87 & 3 \\
PRD168 & 19 49 56.99 & -51 44 55.8 & 17.99 & -0.0004 & 4.62 & 3 \\
PRD169 & 19 50 58.36 & -52 01 31.3 & 18.86 & 0.3263 & 3.38 & 4 \\
PRD170 & 19 49 21.80 & -51 53 39.0 & 18.40 & 0.0002 & 7.31 & 4 \\
PRD171 & 19 49 20.35 & -51 52 8.2 & 10.51 & 0.0433 & 4.95 & 3 \\
PRD172 & 19 52 4.71 & -51 59 22.0 & 18.04 & 0.1056 & 5.69 & 3 & yes \\
PRD173 & 19 52 13.94 & -52 03 2.8 & 16.48 & 0.0000 & 4.74 & 3 \\
PRD174 & 19 49 42.22 & -52 02 2.0 & 18.78 & 0.0527 & 2.94 & 4 \\
PRD175 & 19 49 8.45 & -52 03 48.5 & 17.10 & 0.0003 & 12.75 & 4 \\
PRD176 & 19 49 35.06 & -52 13 36.9 & 18.30 & 0.0002 & 4.33 & 3 \\
PRD177 & 19 52 11.96 & -52 00 41.3 & 17.69 & 0.0993 & 3.60 & 4 & yes \\
PRD178 & 19 49 36.04 & -52 10 24.0 & 18.76 & 0.0008 & 8.29 & 4 \\
PRD179 & 19 49 27.10 & -52 24 25.4 & 17.09 & 0.0581 & 3.59 & 4 \\
PRD180 & 19 49 48.77 & -52 25 15.3 & 19.14 & 0.0509 & 3.25 & 4 \\
PRD181 & 19 51 52.34 & -52 14 30.0 & 18.60 & 0.1059 & 6.05 & 3 & yes \\
PRD182 & 19 52 14.69 & -52 18 37.4 & 19.25 & 0.1573 & 2.95 & 4 \\
PRD183 & 19 52 9.11 & -52 26 38.1 & 17.28 & 0.0004 & 3.91 & 3 \\
PRD184 & 19 51 19.93 & -52 11 17.9 & 18.20 & 0.0008 & 3.64 & 3 \\
PRD185 & 19 53 28.70 & -52 20 7.1 & 18.79 & 0.1913 & 3.33 & 4 \\
PRD186 & 19 53 27.48 & -52 13 41.3 & 18.26 & 0.1079 & 5.83 & 3 & yes \\
\hline
\noalign{\smallskip}
\end{tabular}
\end{center}
\end{table*}

%
%
\begin{table*}
\setcounter{table}{2}
\begin{center}
\caption{continued.
\hfil}
\begin{tabular}{lllccccl}
\noalign{\medskip}
\hline
Identification & RA & Dec & R & z & TDV & Quality & Abell~3653 \\
Tag            & (J2000) & (J2000) & &  & &      & Member? \\
\hline
PRD187 & 19 54 4.78 & -52 15 41.5 & 18.90 & 0.1059 & 4.65 & 3 & yes \\
PRD188 & 19 54 31.16 & -52 33 6.5 & 19.06 & 0.0489 & 3.64 & 4 \\
PRD189 & 19 55 23.48 & -52 33 3.4 & 19.07 & 0.0911 & 4.17 & 3 \\
PRD190 & 19 55 23.18 & -52 27 57.1 & 16.39 & 0.0001 & 13.85 & 4 \\
PRD191 & 19 55 13.49 & -52 28 34.3 & 14.46 & 0.0499 & 3.02 & 4 \\
PRD192 & 19 53 26.21 & -52 08 0.3 & 19.13 & 0.1147 & 4.03 & 4 & yes \\
PRD193 & 19 55 39.76 & -52 19 12.8 & 17.43 & 0.0003 & 4.91 & 3 \\
PRD194 & 19 54 35.14 & -52 11 37.9 & 19.07 & 0.0002 & 4.41 & 3 \\
PRD195 & 19 55 55.08 & -52 09 16.3 & 18.59 & -0.0001 & 7.41 & 4 \\
PRD196 & 19 53 3.52 & -52 02 16.2 & 11.21 & 0.1099 & 5.39 & 4 & yes \\
PRD197 & 19 55 3.24 & -52 07 11.8 & 18.86 & 0.1075 & 4.13 & 4 & yes \\
PRD198 & 19 53 0.46 & -52 09 5.1 & 18.85 & 0.1053 & 3.71 & 4 & yes \\
PRD199 & 19 53 19.98 & -52 10 47.6 & 18.92 & -0.0001 & 5.24 & 3 \\
PRD200 & 19 55 50.30 & -51 55 2.4 & 15.38 & 0.0001 & 10.36 & 4 \\
PRD201 & 19 55 37.48 & -51 54 36.5 & 15.98 & 0.0000 & 10.78 & 4 \\
PRD202 & 19 52 46.70 & -52 00 58.0 & 17.91 & 0.0000 & 8.30 & 4 \\
PRD203 & 19 53 31.84 & -51 54 6.3 & 17.28 & -0.0003 & 11.53 & 4 \\
PRD204 & 19 56 1.91 & -51 44 24.4 & 17.82 & -0.0001 & 7.70 & 4 \\
PRD205 & 19 55 36.43 & -51 42 18.0 & 17.62 & -0.0003 & 8.45 & 4 \\
PRD206 & 19 54 28.55 & -51 48 26.9 & 17.98 & 0.0003 & 5.12 & 3 \\
PRD207 & 19 55 38.83 & -51 43 53.0 & 18.38 & 0.0003 & 5.16 & 3 \\
PRD208 & 19 55 34.74 & -51 40 4.6 & 18.17 & -0.0004 & 6.16 & 3 \\
PRD209 & 19 55 46.20 & -51 40 23.5 & 16.94 & 0.0000 & 6.85 & 3 \\
PRD210 & 19 55 45.48 & -51 31 25.0 & 18.48 & -0.0004 & 4.12 & 3 \\
PRD211 & 19 55 35.36 & -51 31 7.6 & 17.88 & 0.0005 & 6.79 & 3 \\
PRD212 & 19 54 54.52 & -51 43 21.1 & 18.60 & 0.0003 & 3.60 & 3 \\
PRD213 & 19 54 18.45 & -51 43 32.8 & 18.46 & 0.0009 & 3.08 & 3 \\
PRD214 & 19 54 36.23 & -51 42 3.3 & 18.67 & 0.0018 & 3.81 & 3 \\
PRD215 & 19 55 15.00 & -51 31 21.2 & 17.84 & 0.0023 & 4.29 & 3 \\
PRD216 & 19 54 13.98 & -51 30 29.0 & 18.45 & -0.0001 & 5.21 & 3 \\
PRD217 & 19 53 46.29 & -51 37 32.6 & 16.75 & -0.0002 & 10.58 & 4 \\
PRD218 & 19 54 22.55 & -51 35 33.6 & 18.56 & 0.0004 & 3.83 & 3 \\
PRD219 & 19 53 48.89 & -51 52 3.3 & 15.34 & -0.0001 & 14.10 & 4 \\
PRD220 & 19 52 54.86 & -51 53 13.2 & 18.16 & 0.0001 & 14.94 & 4 \\
PRD221 & 19 54 20.27 & -51 41 23.3 & 17.71 & -0.0004 & 4.56 & 3 \\
PRD222 & 19 53 12.77 & -51 29 17.0 & 18.65 & 0.1516 & 4.78 & 3 \\
PRD223 & 19 53 7.98 & -51 31 27.6 & 17.83 & 0.0000 & 6.49 & 3 \\
PRD224 & 19 52 58.84 & -51 44 38.5 & 18.49 & -0.0004 & 5.64 & 3 \\
PRD225 & 19 52 52.38 & -51 42 40.5 & 18.14 & 0.0003 & 5.73 & 3 \\
PRD226 & 19 52 49.82 & -51 38 16.7 & 18.44 & 0.0011 & 4.53 & 4 \\
PRD227 & 19 52 44.01 & -51 41 59.3 & 14.37 & 0.0002 & 11.44 & 3 \\
PRD228 & 19 52 0.38 & -51 52 5.8 & 17.74 & -0.0004 & 8.64 & 3 \\
PRD229 & 19 51 37.01 & -51 43 3.6 & 18.85 & 0.0003 & 6.54 & 3 \\
PRD230 & 19 52 14.67 & -52 00 29.1 & 17.86 & 0.1074 & 7.62 & 3 & yes \\
PRD231 & 19 52 23.85 & -52 00 32.3 & 18.09 & -0.0005 & 4.06 & 3 \\
PRD232 & 19 51 58.20 & -51 36 53.8 & 15.85 & -0.0001 & 14.32 & 3 \\
PRD233 & 19 52 9.20 & -51 46 1.1 & 16.67 & 0.0005 & 4.77 & 3 \\
PRD234 & 19 51 53.13 & -51 49 40.6 & 18.48 & -0.0001 & 9.69 & 3 \\
PRD235 & 19 50 42.36 & -51 32 52.2 & 16.95 & -0.0003 & 9.63 & 4 \\
PRD236 & 19 50 16.06 & -51 29 43.8 & 16.83 & -0.0002 & 7.41 & 4 \\
PRD237 & 19 51 9.95 & -51 44 25.5 & 16.95 & 0.0001 & 5.72 & 3 \\
PRD238 & 19 49 25.53 & -51 35 33.9 & 18.32 & 0.0000 & 5.25 & 3 \\
PRD239 & 19 51 22.96 & -51 54 54.8 & 16.71 & -0.0003 & 11.31 & 4 \\
PRD240 & 19 50 43.09 & -51 51 7.0 & 18.03 & -0.0002 & 7.56 & 4 \\
PRD241 & 19 51 31.23 & -51 53 38.5 & 17.73 & 0.0000 & 11.10 & 4 \\
PRD242 & 19 49 59.36 & -51 43 1.7 & 17.82 & 0.0003 & 9.99 & 4 \\
PRD243 & 19 50 6.28 & -51 40 12.1 & 17.91 & 0.0000 & 4.35 & 3 \\
PRD244 & 19 49 19.09 & -51 44 13.8 & 18.27 & -0.0004 & 3.39 & 3 \\
PRD245 & 19 52 19.61 & -52 01 18.9 & 17.65 & 0.1054 & 5.34 & 3 & yes \\
PRD246 & 19 49 9.94 & -51 59 17.5 & 17.68 & 0.0000 & 5.23 & 3 \\
PRD247 & 19 50 47.63 & -52 12 28.9 & 17.29 & 0.0003 & 5.83 & 3 \\
PRD248 & 19 50 57.12 & -52 13 2.9 & 18.20 & -0.0002 & 6.63 & 3 \\
PRD249 & 19 50 32.31 & -52 05 11.9 & 15.98 & -0.0002 & 13.21 & 4 \\
\hline
\noalign{\smallskip}
\end{tabular}
\end{center}
\end{table*}

%
%
\begin{table*}
\setcounter{table}{2}
\begin{center}
\caption{continued.
\hfil}
\begin{tabular}{lllccccl}
\noalign{\medskip}
\hline
Identification & RA & Dec & R & z & TDV & Quality & Abell~3653 \\
Tag            & (J2000) & (J2000) & &  & &      & Member? \\
\hline
PRD250 & 19 51 30.41 & -52 11 1.8 & 15.09 & 0.0001 & 17.37 & 4 \\
PRD251 & 19 50 12.91 & -52 07 9.6 & 17.06 & 0.0001 & 5.66 & 3 \\
PRD252 & 19 49 53.77 & -52 10 11.8 & 18.19 & -0.0005 & 4.18 & 3 \\
PRD253 & 19 49 36.68 & -52 15 31.3 & 16.28 & 0.0001 & 16.95 & 4 \\
PRD254 & 19 49 13.63 & -52 18 18.4 & 17.28 & 0.0001 & 7.26 & 4 \\
PRD255 & 19 49 29.20 & -52 18 56.4 & 18.47 & 0.0000 & 6.18 & 3 \\
PRD256 & 19 49 18.89 & -52 21 0.8 & 17.72 & 0.0000 & 5.65 & 3 \\
PRD257 & 19 50 22.82 & -52 18 2.4 & 18.09 & -0.0003 & 5.57 & 3 \\
PRD258 & 19 49 30.12 & -52 24 33.0 & 16.25 & 0.0001 & 20.81 & 4 \\
PRD259 & 19 51 4.87 & -52 15 40.2 & 17.97 & 0.0005 & 5.79 & 3 \\
PRD260 & 19 49 22.48 & -52 31 48.9 & 15.83 & -0.0001 & 9.57 & 3 \\
PRD261 & 19 49 48.67 & -52 30 3.1 & 18.69 & -0.0008 & 5.09 & 3 \\
PRD262 & 19 51 41.58 & -52 14 29.5 & 16.07 & 0.0000 & 11.79 & 4 \\
PRD263 & 19 50 12.21 & -52 26 25.5 & 18.20 & -0.0001 & 5.80 & 3 \\
PRD264 & 19 50 34.38 & -52 22 40.3 & 16.25 & -0.0002 & 14.37 & 4 \\
PRD265 & 19 51 0.59 & -52 30 11.0 & 17.52 & -0.0001 & 8.91 & 4 \\
PRD266 & 19 52 12.60 & -52 04 7.3 & 16.08 & -0.0003 & 10.70 & 4 \\
PRD267 & 19 51 21.74 & -52 30 39.1 & 15.84 & -0.0001 & 8.74 & 4 \\
PRD268 & 19 52 13.58 & -52 24 24.4 & 17.76 & 0.0000 & 4.47 & 3 \\
PRD269 & 19 52 9.08 & -52 32 20.6 & 14.89 & 0.0011 & 4.76 & 3 \\
PRD270 & 19 51 53.23 & -52 19 18.8 & 18.34 & -0.0002 & 3.79 & 3 \\
PRD271 & 19 52 34.26 & -52 07 31.1 & 17.12 & -0.0001 & 6.75 & 3 \\
PRD272 & 19 52 22.27 & -52 28 48.3 & 15.03 & 0.0001 & 12.46 & 4 \\
PRD273 & 19 53 15.21 & -52 31 29.6 & 18.20 & 0.0000 & 10.58 & 4 \\
PRD274 & 19 53 7.61 & -52 21 56.6 & 15.88 & -0.0001 & 14.15 & 4 \\
PRD275 & 19 53 22.47 & -52 27 19.2 & 12.16 & -0.0001 & 11.67 & 4 \\
PRD276 & 19 53 35.45 & -52 32 33.4 & 18.81 & 0.0001 & 24.29 & 4 \\
PRD277 & 19 53 4.62 & -52 25 19.2 & 16.55 & -0.0001 & 13.98 & 4 \\
PRD278 & 19 53 13.00 & -52 17 45.4 & 16.46 & -0.0003 & 11.42 & 4 \\
PRD279 & 19 53 58.33 & -52 04 10.8 & 18.27 & 0.1090 & 3.76 & 3 & yes \\
PRD280 & 19 53 6.55 & -52 07 51.4 & 18.76 & 0.0002 & 3.80 & 3 \\
PRD281 & 19 54 40.88 & -52 19 58.0 & 18.24 & 0.0004 & 5.31 & 3 \\
PRD282 & 19 53 47.75 & -52 04 28.7 & 18.04 & 0.1092 & 5.32 & 3 & yes \\
PRD283 & 19 53 58.70 & -52 18 16.9 & 18.25 & 0.0003 & 4.46 & 3 \\
PRD284 & 19 54 51.14 & -52 23 10.4 & 18.66 & -0.0006 & 3.24 & 3 \\
PRD285 & 19 53 6.55 & -52 02 20.6 & 18.14 & 0.1099 & 5.44 & 3 & yes \\
PRD286 & 19 55 41.81 & -52 28 34.0 & 17.86 & -0.0001 & 8.47 & 3 \\
PRD287 & 19 55 41.77 & -52 26 8.5 & 18.73 & -0.0007 & 4.31 & 3 \\
PRD288 & 19 55 42.43 & -52 21 39.4 & 16.91 & -0.0001 & 11.41 & 4 \\
PRD289 & 19 53 42.04 & -52 10 5.3 & 16.39 & -0.0003 & 16.20 & 4 \\
PRD290 & 19 56 6.09 & -52 16 29.5 & 17.60 & -0.0001 & 3.43 & 3 \\
PRD291 & 19 56 6.30 & -52 04 36.5 & 17.13 & -0.0005 & 10.55 & 4 \\
PRD292 & 19 55 30.82 & -52 03 16.0 & 18.74 & 0.0000 & 3.76 & 3 \\
PRD293 & 19 55 28.03 & -52 01 6.1 & 17.49 & 0.0007 & 6.89 & 4 \\
PRD294 & 19 49 16.64 & -51 40 57.8 & 17.98 & -0.0003 & 5.38 & 3 \\
PRD295 & 19 49 18.31 & -51 33 28.9 & 18.30 & 0.0281 & 4.41 & 4 \\
PRD296 & 19 49 21.12 & -52 26 12.2 & 18.74 & 0.0953 & 3.23 & 3 \\
PRD297 & 19 49 24.05 & -52 08 50.1 & 17.44 & 0.0942 & 8.63 & 4 \\
PRD298 & 19 49 24.70 & -52 32 40.8 & 18.29 & 0.3061 & 4.32 & 3 \\
PRD299 & 19 49 25.32 & -52 21 48.4 & 16.60 & -0.0001 & 5.94 & 4 \\
PRD300 & 19 49 28.80 & -51 50 21.8 & 18.82 & 0.0000 & 4.46 & 3 \\
PRD301 & 19 49 53.20 & -52 26 7.6 & 18.36 & 0.0507 & 3.01 & 4 \\
PRD302 & 19 49 55.86 & -52 05 47.4 & 17.18 & -0.0001 & 6.82 & 3 \\
PRD303 & 19 50 11.59 & -51 48 23.4 & 17.87 & 0.0004 & 6.89 & 4 \\
PRD304 & 19 50 19.45 & -52 01 30.0 & 17.44 & 0.1040 & 6.05 & 3 & yes \\
PRD305 & 19 50 21.46 & -51 41 40.6 & 15.84 & -0.0002 & 5.13 & 3 \\
PRD306 & 19 50 27.10 & -51 55 39.1 & 17.02 & 0.0735 & 6.64 & 3 \\
PRD307 & 19 50 27.50 & -51 37 10.0 & 16.35 & 0.0567 & 5.67 & 4 \\
PRD308 & 19 50 31.74 & -52 32 36.5 & 18.11 & 0.0954 & 5.11 & 3 \\
PRD309 & 19 50 42.45 & -51 38 41.9 & 15.93 & 0.1068 & 7.31 & 4 & yes \\
PRD310 & 19 50 43.52 & -51 52 21.3 & 17.72 & 0.1067 & 7.18 & 4 & yes \\
PRD311 & 19 50 4.70 & -51 51 31.2 & 17.39 & 0.0572 & 4.70 & 4 \\
PRD312 & 19 50 5.12 & -51 37 48.4 & 18.79 & 0.2494 & 3.37 & 4 \\
\hline
\noalign{\smallskip}
\end{tabular}
\end{center}
\end{table*}

%
%
\begin{table*}
\setcounter{table}{2}
\begin{center}
\caption{continued.
\hfil}
\begin{tabular}{lllccccl}
\noalign{\medskip}
\hline
Identification & RA & Dec & R & z & TDV & Quality & Abell~3653 \\
Tag            & (J2000) & (J2000) & &  & &      & Member? \\
\hline
PRD313 & 19 51 10.23 & -52 01 55.3 & 17.69 & 0.1577 & 7.70 & 4 \\
PRD314 & 19 51 13.61 & -52 20 25.9 & 16.40 & 0.0003 & 9.70 & 4 \\
PRD315 & 19 51 15.59 & -51 49 3.9 & 17.49 & 0.1064 & 9.71 & 4 & yes \\
PRD316 & 19 51 16.72 & -52 06 18.5 & 18.72 & 0.1080 & 3.47 & 4 & yes \\
PRD317 & 19 51 18.34 & -51 58 58.6 & 18.86 & 0.0509 & 3.01 & 3 \\
PRD318 & 19 51 20.86 & -52 19 41.2 & 17.79 & 0.0000 & 5.54 & 3 \\
PRD319 & 19 51 21.85 & -51 47 13.3 & 17.33 & 0.0579 & 4.94 & 4 \\
PRD320 & 19 51 2.67 & -52 26 49.3 & 18.99 & 0.0489 & 4.26 & 4 \\
PRD321 & 19 51 30.99 & -52 26 47.5 & 16.67 & -0.0001 & 10.61 & 4 \\
PRD322 & 19 51 32.39 & -51 56 32.6 & 14.92 & 0.0392 & 3.51 & 4 \\
PRD323 & 19 51 33.12 & -52 02 55.5 & 16.93 & 0.1074 & 8.69 & 4 & yes \\
PRD324 & 19 51 39.34 & -52 02 31.0 & 17.22 & 0.1097 & 3.71 & 4 & yes \\
PRD325 & 19 51 39.83 & -52 08 33.9 & 18.45 & 0.1068 & 6.24 & 3 & yes \\
PRD326 & 19 51 41.80 & -52 26 58.2 & 18.43 & 0.0007 & 5.39 & 4 \\
PRD327 & 19 51 4.57 & -51 33 51.6 & 19.17 & 0.1089 & 5.31 & 3 & yes \\
PRD328 & 19 51 46.97 & -51 58 18.0 & 18.43 & 0.1079 & 6.81 & 4 & yes \\
PRD329 & 19 51 49.76 & -52 33 29.2 & 18.13 & 0.1075 & 3.80 & 3 & yes \\
PRD330 & 19 51 50.19 & -51 48 4.3 & 17.47 & 0.1021 & 7.95 & 4 & yes \\
PRD331 & 19 51 54.26 & -51 57 52.0 & 17.76 & 0.1084 & 7.06 & 4 & yes \\
PRD332 & 19 51 57.69 & -52 00 25.9 & 18.32 & 0.1048 & 6.62 & 3 & yes \\
PRD333 & 19 51 58.55 & -52 03 8.4 & 16.21 & 0.1093 & 5.90 & 4 & yes \\
PRD334 & 19 52 11.21 & -51 58 1.5 & 17.98 & 0.1040 & 4.68 & 3 & yes \\
PRD335 & 19 52 12.80 & -52 07 14.8 & 16.76 & 0.0000 & 7.16 & 3 \\
PRD336 & 19 52 14.09 & -51 32 6.8 & 17.51 & 0.1076 & 7.15 & 4 & yes \\
PRD337 & 19 52 16.36 & -51 43 51.6 & 18.39 & 0.0939 & 3.87 & 4 \\
PRD338 & 19 52 29.65 & -52 05 7.4 & 17.71 & 0.1063 & 7.85 & 3 & yes \\
PRD339 & 19 52 33.23 & -52 04 16.7 & 18.39 & 0.0944 & 3.91 & 3 \\
PRD340 & 19 52 33.84 & -51 55 1.9 & 18.17 & 0.1083 & 7.79 & 3 & yes \\
PRD341 & 19 52 35.11 & -51 39 31.6 & 17.76 & 0.1069 & 7.55 & 3 & yes \\
PRD342 & 19 52 38.59 & -51 29 43.9 & 18.97 & 0.1336 & 3.04 & 4 \\
PRD343 & 19 52 43.30 & -51 50 20.8 & 18.58 & 0.0564 & 3.48 & 4 \\
PRD344 & 19 52 44.32 & -52 16 10.4 & 17.73 & 0.0002 & 4.13 & 3 \\
PRD345 & 19 52 49.98 & -52 01 41.2 & 17.53 & 0.1065 & 3.16 & 3 & yes \\
PRD346 & 19 52 54.44 & -51 51 25.6 & 18.53 & 0.1024 & 3.09 & 4 & yes \\
PRD347 & 19 52 54.45 & -52 03 26.4 & 17.46 & 0.1092 & 6.93 & 3 & yes \\
PRD348 & 19 52 57.09 & -52 02 45.7 & 16.27 & 0.1097 & 6.75 & 3 & yes \\
PRD349 & 19 52 57.49 & -51 58 17.1 & 17.91 & 0.1055 & 8.31 & 3 & yes \\
PRD350 & 19 52 9.52 & -52 03 39.9 & 17.39 & 0.1056 & 4.13 & 4 & yes \\
PRD351 & 19 53 0.70 & -52 01 14.3 & 17.79 & 0.1087 & 7.51 & 3 & yes \\
PRD352 & 19 53 1.00 & -51 48 35.3 & 17.60 & 0.1124 & 6.96 & 4 & yes \\
PRD353 & 19 53 12.41 & -52 06 31.6 & 16.91 & 0.1040 & 9.54 & 4 & yes \\
PRD354 & 19 53 16.66 & -51 47 38.5 & 18.71 & -0.0002 & 8.33 & 4 \\
PRD355 & 19 53 22.60 & -52 13 30.0 & 16.52 & 0.1087 & 3.83 & 4 & yes \\
PRD356 & 19 53 2.80 & -51 42 40.8 & 18.83 & 0.1053 & 6.36 & 3 & yes \\
PRD357 & 19 53 29.80 & -52 16 22.2 & 17.02 & 0.1093 & 6.01 & 3  & yes\\
PRD358 & 19 53 33.09 & -51 58 33.1 & 17.81 & 0.1083 & 6.37 & 4 & yes \\
PRD359 & 19 53 35.99 & -51 57 45.1 & 17.99 & 0.1010 & 3.39 & 4 & yes \\
PRD360 & 19 53 36.55 & -52 02 35.9 & 17.81 & 0.0499 & 3.68 & 4 \\
PRD361 & 19 53 3.89 & -52 25 36.5 & 18.44 & 0.1897 & 8.70 & 3 \\
PRD362 & 19 53 40.31 & -52 04 7.7 & 17.88 & 0.1125 & 6.16 & 3 & yes \\
PRD363 & 19 53 40.64 & -52 29 22.6 & 18.44 & 0.1898 & 5.08 & 4 \\
PRD364 & 19 53 42.77 & -52 06 60.0 & 17.29 & 0.1072 & 6.64 & 3 & yes \\
PRD365 & 19 53 4.63 & -51 52 9.3 & 17.47 & 0.1065 & 3.05 & 4 & yes \\
PRD366 & 19 53 55.88 & -52 05 19.4 & 16.41 & 0.1045 & 7.20 & 3 & yes \\
PRD367 & 19 53 8.19 & -51 35 29.9 & 16.29 & 0.0000 & 10.63 & 4 \\
PRD368 & 19 53 9.43 & -52 11 12.0 & 16.77 & 0.1107 & 6.81 & 3 & yes \\
PRD369 & 19 54 19.84 & -52 21 37.8 & 17.22 & 0.1093 & 6.32 & 3 & yes \\
PRD370 & 19 54 19.91 & -51 49 4.0 & 17.36 & 0.0876 & 3.41 & 4 \\
PRD371 & 19 54 22.38 & -52 00 24.5 & 16.69 & 0.1008 & 4.79 & 4 & yes \\
PRD372 & 19 54 30.71 & -52 09 13.4 & 16.54 & 0.1062 & 7.54 & 3 & yes \\
PRD373 & 19 54 35.35 & -52 16 0.9 & 18.34 & 0.1075 & 5.83 & 3 & yes \\
PRD374 & 19 54 3.95 & -51 47 4.9 & 16.99 & -0.0001 & 4.69 & 3 \\
PRD375 & 19 54 42.34 & -51 35 58.0 & 17.84 & -0.0002 & 9.01 & 4 \\
\hline
\noalign{\smallskip}
\end{tabular}
\end{center}
\end{table*}

%
%
\begin{table*}
\setcounter{table}{2}
\begin{center}
\caption{continued.
\hfil}
\begin{tabular}{lllccccl}
\noalign{\medskip}
\hline
Identification & RA & Dec & R & z & TDV & Quality & Abell~3653 \\
Tag            & (J2000) & (J2000) & &  & &      & Member? \\
\hline
PRD376 & 19 54 45.01 & -52 13 43.4 & 18.38 & 0.1458 & 5.20 & 4 \\
PRD377 & 19 54 47.49 & -52 11 54.9 & 17.09 & 0.1092 & 6.40 & 4 & yes \\
PRD378 & 19 54 57.52 & -51 41 25.9 & 17.02 & -0.0005 & 5.58 & 3 \\
PRD379 & 19 54 5.85 & -51 50 44.4 & 18.21 & 0.0507 & 3.45 & 4 \\
PRD380 & 19 54 7.20 & -51 47 34.6 & 19.06 & 0.1149 & 5.52 & 4 & yes \\
PRD381 & 19 55 17.08 & -51 37 32.2 & 17.15 & 0.0580 & 4.32 & 3 \\
PRD382 & 19 55 2.04 & -52 24 42.4 & 16.87 & 0.0573 & 6.22 & 3 \\
PRD383 & 19 55 27.09 & -51 58 53.0 & 18.46 & 0.1558 & 8.32 & 3 \\
PRD384 & 19 55 31.27 & -52 28 3.7 & 17.81 & 0.0002 & 6.72 & 4 \\
PRD385 & 19 55 32.27 & -51 38 14.6 & 17.42 & 0.0868 & 5.96 & 3 \\
PRD386 & 19 55 32.72 & -52 17 16.4 & 18.67 & 0.1997 & 7.64 & 3 \\
PRD387 & 19 55 38.38 & -51 57 4.7 & 18.45 & 0.2279 & 5.46 & 4 \\
PRD388 & 19 55 41.62 & -51 39 20.8 & 17.88 & 0.0880 & 3.30 & 4 \\
PRD389 & 19 55 41.65 & -52 14 49.3 & 18.46 & 0.0995 & 2.75 & 4 & yes \\
PRD390 & 19 55 50.90 & -52 29 52.1 & 16.63 & 0.0499 & 2.90 & 4 \\
PRD391 & 19 55 52.54 & -52 03 8.9 & 16.92 & 0.1078 & 6.03 & 3 & yes \\
\hline
\noalign{\smallskip}
\end{tabular}
  \label{tab:cat}
\end{center}
\end{table*}

\end{document}